\documentclass[aps,prb,reprint,longbibliography,floatfix]{revtex4-2}

\usepackage{amsmath,amssymb,amsfonts,bm,mathtools}
\usepackage{graphicx}
\usepackage{physics}
\usepackage{hyperref}
\usepackage{xcolor}

\hypersetup{colorlinks = true, 
		linkcolor = blue, 
		urlcolor = blue,
		citecolor = blue}

\newcommand{\ii}{\mathrm{i}}
\newcommand{\ee}{\mathrm{e}}
\newcommand{\SL}{S_{N/2}}
\newcommand{\etaloc}{\eta}
\newcommand{\IPR}{\mathrm{IPR}}

\begin{document}

\title{Graded hopping screens nonreciprocity and reorganizes Stark asymptotics in a non-Hermitian Stark chain}

\author{Y. S. Liu}
\affiliation{College of Physics and Materials Science, Tianjin Normal
	University, Tianjin 300387, China}

\author{X. Z. Zhang} \altaffiliation{zhangxz@tjnu.edu.cn}
\affiliation{College of Physics and Materials Science, Tianjin Normal
	University, Tianjin 300387, China}

\begin{abstract}

We study a one-dimensional non-Hermitian Stark chain in which nonreciprocal hopping, a linear potential, and linearly graded hopping act simultaneously. The central question is how boundary pumping and field-induced confinement are reorganized when the hopping amplitude itself grows with position. We show that the graded term separates the two localization channels at the level of the large-position asymptotics. An exact diagonal similarity transformation removes the bond asymmetry and converts the usual exponential skin factor into an algebraic boundary accumulation with exponent $\eta=\gamma/F_2$. The transformed symmetric chain then reduces asymptotically to a constant-coefficient recurrence, giving the Stark threshold $|F_1|=2|F_2|$. The original right eigenstates acquire the unified envelope $\psi_j^R\sim j^{\eta}\phi_j$, with oscillatory, double-root, and exponentially localized branches across the threshold. This form also yields two finite-size scales, one measuring the logarithmic screening of nonreciprocity and the other balancing the algebraic skin factor against the exponential Stark tail. A joint localization map in the $(\gamma,F_1/F_2)$ plane verifies this structure. The edge polarization bends near the Stark threshold and weakens on the localized side, while the inverse participation ratio of the most localized eigenstates rises rapidly for $F_1/F_2>2$. Using a normalized Gaussian projector appropriate for non-unitary evolution, we further show that the same threshold enhances half-chain entanglement growth after a charge-density-wave quench. These results identify graded hopping as a controlled mechanism for screening nonreciprocity, resetting Stark asymptotics, and organizing the finite-size crossover between algebraic skin accumulation and Stark localization.

\end{abstract}

\maketitle

\section{Introduction}

Non-Hermitian band theory has made boundary sensitivity a central issue in quantum and wave physics. The most direct manifestation is the non-Hermitian skin effect, in which an extensive number of open-boundary eigenstates accumulate at one edge of the system \cite{HatanoNelson1996,YokomizoMurakami2019,Okuma2020,PhysRevLett.123.066404,PhysRevLett.121.086803,Xiao2020}. This accumulation is not a small boundary correction. It changes bulk-boundary correspondence, encodes point-gap spectral topology, and affects wave-packet dynamics, transport, and entanglement diagnostics \cite{Okuma2020,Kawabata2023}. It has also become experimentally accessible in engineered topolectrical, acoustic, photonic, and mechanical platforms \cite{Zhang2021HigherOrderNHSE,Li2024DynamicNHSE}. These developments make it important to understand when skin accumulation is robust, when it is screened, and how it competes with other mechanisms of localization.

A linear potential provides a second and conceptually distinct route to localization. In Hermitian lattices it produces Wannier-Stark ladders, Bloch oscillations, and field-induced confinement without relying on disorder \cite{BenDahan1996,Wilkinson1996,PhysRevA.61.061402}. Recent experiments and simulations have made this physics directly visible in cold-atom and superconducting quantum platforms, where gradients, initial states, and dynamical observables can be controlled with high precision \cite{Guo2021WSL,Karamlou2022}. In non-Hermitian settings, a Stark field is especially useful because it competes with boundary pumping in a deterministic way. It can suppress, reshape, or dynamically control skin accumulation, but the result depends sensitively on how the hopping scale enters the large-distance problem \cite{PhysRevB.106.L161402,PhysRevB.110.094310,PhysRevB.111.174202,Zheng2024}.

This observation motivates the present work. We ask what happens when nonreciprocal hopping, a linear potential, and linearly graded hopping are present in the same open chain. The graded hopping is not introduced as a weak spatial perturbation. At large position, the hopping term $jF_2$ grows in the same way as the Stark potential $F_1j$, while the bare nonreciprocal imbalance remains of order one. Thus the relative nonreciprocal bias decreases as $1/j$, whereas the ratio between the Stark slope and the graded hopping slope remains finite. This simple scaling argument already suggests that graded hopping should screen the skin effect while simultaneously setting the Stark threshold. The model is therefore a minimal setting in which boundary pumping, field-induced confinement, and spatially growing kinetic energy can be separated analytically.

The main analytical structure follows from two steps. First, a diagonal similarity transformation removes the bond-by-bond asymmetry. In a uniform nonreciprocal chain this procedure produces an exponential gauge factor. In the present graded chain the local gauge increment decays as $1/j$, so the accumulated factor is algebraic. Second, the transformed chain has a large-position recurrence in which both the hopping and the potential are proportional to $j$. Dividing by $j$ gives a constant-coefficient recurrence and the asymptotic threshold
\begin{equation}
|F_1|=2|F_2|.
\label{eq:main_threshold}
\end{equation}
Combining the two pieces gives the original right eigenstate the unified asymptotic envelope
\begin{equation}
\psi_j^R\sim j^{\eta}\phi_j,
\label{eq:unified_envelope_intro}
\end{equation}
where $j^{\eta}$ is the residual algebraic skin factor and $\phi_j$ is fixed by the Stark recurrence. The three localization mechanisms are therefore not treated as independent effects. They enter one asymptotic expression.

This analytical split also fixes the numerical strategy. The residual skin sector is controlled by a finite-size screening variable proportional to $(\gamma/F_2)\ln N$. The Stark-localized side is controlled by a second scale obtained by balancing the algebraic factor against the exponential Stark tail. We therefore use one-body observables that separately track boundary polarization and localization strength, and we use the many-body dynamics only after the one-body structure has been identified. A technical point is essential for this comparison. Under non-unitary evolution, occupied orbitals of a Slater determinant do not remain orthonormal. The correlation matrix must therefore be built from the normalized Gaussian projector rather than from the naive product of evolved orbitals. This convention gives the entanglement entropy of the normalized right-state evolution used throughout this work and avoids mixing it with a biorthogonal diagnostic.

The remainder of the paper is organized as follows. Section~\ref{Model_and_diagnostics} introduces the model and the properly normalized diagnostics. Section~\ref{Exact_similarity_transformation_and_algebraic_skin_factor} derives the exact similarity transformation and the algebraic skin factor. Section~\ref{Unified asymptotic envelopes and finite-size competition scales} derives the unified asymptotic envelopes and the associated finite-size competition scales. Section~\ref{Joint localization map} presents the joint localization map implied by the analytics. Section~\ref{Many-body entanglement growth} shows that the same threshold leaves a clear dynamical trace in the growth of the half-chain many-body entanglement entropy. Section~\ref{Discussion and conclusion} closes with a discussion and a summary of the physical picture.

\section{Model and diagnostics}
\label{Model_and_diagnostics}
We consider the open-chain quadratic Hamiltonian
\begin{align}
H &= \sum_{j=1}^{N-1}\left[(J-\gamma+jF_2)c_j^\dagger c_{j+1}+(J+\gamma+jF_2)c_{j+1}^\dagger c_j\right]
\nonumber\\
&\quad + F_1\sum_{j=1}^N j\, c_j^\dagger c_j.
\label{eq:H_manybody}
\end{align}
In the one-body basis $\ket{j}=c_j^\dagger\ket{0}$, the Hamiltonian reads
\begin{equation}
	H = \sum_{j=1}^{N-1}\left(t_j^L\ket{j}\bra{j+1}+t_j^R\ket{j+1}\bra{j}\right)
	+\sum_{j=1}^N F_1 j\ket{j}\bra{j},
	\label{eq:H_single}
\end{equation}
with
\begin{equation}
	t_j^L = J-\gamma+jF_2,
	\qquad
	t_j^R = J+\gamma+jF_2.
	\label{eq:graded_hoppings}
\end{equation}
The chain contains $N$ lattice sites, labeled by $j=1,2,\dots,N$, and open boundaries are imposed throughout. The parameter $\gamma$ measures nonreciprocity, $F_1$ is the Stark slope, and $F_2$ is the hopping gradient. We take $F_2>0$ in the numerical work. The parameter $J$ fixes the uniform offset of the hopping. The choice $J=1$ is used in the one-body localization maps, while $J=-1$ is used only in the Hermitian dynamical benchmark. At finite $F_2$ these two choices should not be regarded as a general unitary equivalence, since the graded term is not globally sign reversed. What is independent of this offset is the large-$j$ asymptotics. The difference $t_j^R-t_j^L=2\gamma$ is independent of $J$, and the leading graded hopping is $jF_2$ in either case. Hence the algebraic skin exponent, the threshold $|F_1|=2|F_2|$, and the competition scales derived below are unchanged by the uniform offset. Energies are measured in units of $|J|$ and times in units of $|J|^{-1}$ when $J\ne0$. If some short-distance bonds have $t_j^Lt_j^R<0$, the similarity factor may acquire a phase and the transformed symmetric hopping may become complex. Such finite-$j$ features do not affect the large-position results. Isolated zeros of $t_j^L$ or $t_j^R$ are special bond-decoupling points and are not used in the one-body data shown below.

Throughout the one-body analysis we focus on normalized right-eigenstate weights
\begin{equation}
\rho_j^{(n)}=\frac{|\psi_{n,j}^R|^2}{\sum_{\ell=1}^N |\psi_{n,\ell}^R|^2},
\label{eq:rhoj}
\end{equation}
which define the center of mass and inverse participation ratio \cite{ghoshExploringTopologicalLocalization2025},
\begin{equation}
X_n = \sum_{j=1}^N \frac{j-1}{N-1}\rho_j^{(n)},
\qquad
\IPR_n = \sum_{j=1}^N \left(\rho_j^{(n)}\right)^2.
\label{eq:single_particle_diagnostics}
\end{equation}
To visualize the boundary bias more directly, we also introduce the edge-polarization variable
\begin{equation}
\mathcal P_n = 2X_n-1.
\label{eq:edge_pol_state}
\end{equation}
This quantity vanishes for a state centered at the middle of the chain and increases as the weight shifts toward the right boundary. For $\gamma>0$ one has $t_j^R>t_j^L$, so positive $\mathcal P_n$ signals the right-edge accumulation expected from the nonreciprocal bias.

For non-Hermitian many-body free fermions, the normalized correlation matrix requires more care than in the Hermitian case. We use the normalized right-state, or post-selected, Gaussian state obtained by evolving the occupied orbitals with the one-body Hamiltonian. Let these orbitals be collected into an $N\times N_p$ matrix $U$, whose columns need not remain orthonormal under non-unitary evolution. The overlap matrix is
\begin{equation}
M=U^\dagger U.
\end{equation}
The projector onto the occupied single-particle subspace of the normalized Slater determinant is
\begin{equation}
P=U M^{-1}U^\dagger.
\end{equation}
With the convention $C_{ij}=\langle c_i^\dagger c_j\rangle$, the correlation matrix is the transpose of this projector,
\begin{equation}
C=P^T=\left(U M^{-1}U^\dagger\right)^T.
\label{eq:normalized_projector}
\end{equation}
If the occupied orbitals are orthonormal, $M=\mathbb I$ and this reduces to the usual Slater determinant result, up to the same convention-dependent transpose. The transpose does not affect the spectrum of the restricted correlation matrix. The construction satisfies $P^2=P$, and it gives the von Neumann entropy of the normalized right-state evolution. We do not use a biorthogonal entropy in this work. In the dynamical setting,
\begin{align}
U(t)&=\ee^{-\ii Ht}U(0),\\
C(t)&=\left\{U(t)\left[U^\dagger(t)U(t)\right]^{-1}U^\dagger(t)\right\}^T.
\label{eq:time_dep_projector}
\end{align}
Restricting $C(t)$ to the left half of the chain, $A=\{1,2,\dots,N/2\}$, gives $C_A(t)$. If $\{\lambda_\alpha(t)\}$ are the eigenvalues of $C_A(t)$, the half-chain entanglement entropy is \cite{PhysRevB.110.094310,PhysRevB.108.165126,liDisorderInducedEntanglementPhase2023a}
\begin{equation}
\SL(t)=-\sum_\alpha\left[\lambda_\alpha\ln\lambda_\alpha+\left(1-\lambda_\alpha\right)\ln\left(1-\lambda_\alpha\right)\right].
\label{eq:entropy_formula}
\end{equation}
This is the expression used below for many-body dynamics.
	
	In the dynamical part of the paper we use the half-filled charge-density-wave product state
	\begin{equation}
		\ket{\Psi_0}=\prod_{m=1}^{N/2}c_{2m}^\dagger\ket{\mathrm{vac}},
		\label{eq:cdw_state}
	\end{equation}
	where $\ket{\mathrm{vac}}$ represents the vacuum state. This state has no initial bipartite entanglement. The subsequent growth of $\SL(t)$ therefore tracks the spreading and mixing generated by the one-body dynamics without any contamination from initial long-range correlations.

\section{Exact similarity transformation and algebraic skin factor}
\label{Exact_similarity_transformation_and_algebraic_skin_factor}

\begin{figure}[t]
\centering
\includegraphics[width=0.92\linewidth]{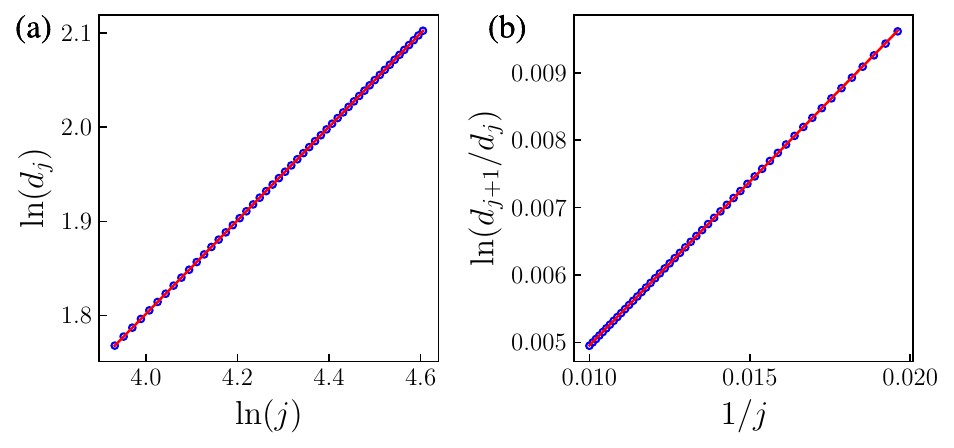}
\caption{Direct verification of the algebraic skin factor for $N=100$, $J=1$, $F_1=0$, $F_2=1$, and $\gamma=0.5$, for which the analytical prediction is $\eta=\gamma/F_2=0.5$. Panel (a) shows the exact diagonal similarity factor $d_j$ versus site index $j$ on a log-log scale together with a power-law fit. In the displayed fitting window the fitted slope is $\eta_{\mathrm{fit}}\simeq0.497$, in good agreement with the analytical value $0.5$. Panel (b) shows the local logarithmic increment $\ln(d_{j+1}/d_j)$ as a function of $1/j$. A linear fit in the asymptotic window gives $\ln(d_{j+1}/d_j)\simeq0.486/j$, again very close to the expected coefficient $\eta=0.5$. According to Eq.~\eqref{eq:local_log_increment}, its asymptotic slope is again fixed by $\eta$. The figure verifies that the ordinary exponential skin accumulation is softened into an algebraic boundary skew once the hopping grows linearly with position.}
\label{fig:algebraic_check}
\end{figure}

For the nonreciprocal Hamiltonian described by Eq.~\eqref{eq:H_single}, its nonreciprocity can be eliminated by a suitable similarity transformation. To this end, we introduce a diagonal similarity transformation,
\begin{equation}
D=\sum_{j=1}^N d_j\ket{j}\bra{j},
\qquad
\widetilde H=D^{-1}HD.
\label{eq:similarity_def}
\end{equation}
The forward hopping on bond $j$ becomes $t_j^L d_{j+1}/d_j$, while the backward hopping becomes $t_j^R d_j/d_{j+1}$. Requiring the two amplitudes to coincide gives
\begin{equation}
\frac{d_{j+1}}{d_j}=\sqrt{\frac{t_j^R}{t_j^L}}=\sqrt{\frac{J+\gamma+jF_2}{J-\gamma+jF_2}}.
\label{eq:s_ratio}
\end{equation}
With this choice the transformed Hamiltonian is
\begin{equation}
\widetilde H = \sum_{j=1}^{N-1}\tau_j\left(\ket{j}\bra{j+1}+\ket{j+1}\bra{j}\right)
+\sum_{j=1}^{N}F_1 j\ket{j}\bra{j},
\label{eq:H_tilde}
\end{equation}
where
\begin{equation}
\tau_j=\sqrt{t_j^Lt_j^R}=\sqrt{(J-\gamma+jF_2)(J+\gamma+jF_2)}.
\label{eq:tauj}
\end{equation}
The bond asymmetry is therefore removed exactly. The transformed chain is symmetric, while the directional bias is carried entirely by the diagonal factor $D$. The transformation is exact whenever no hopping coefficient on the bonds under consideration vanishes. If a zero hopping appears, the chain is split at that bond and the corresponding point should be treated as a separate finite-size decoupling case. In the parameter regime used for Fig.~\ref{fig:algebraic_check} and Fig.~\ref{fig:joint_map}, all products $t_j^Lt_j^R$ are positive, so $D$ may be chosen real and positive.

Iterating Eq.~\eqref{eq:s_ratio} gives
\begin{equation}
d_j=d_1\prod_{m=1}^{j-1}\sqrt{\frac{J+\gamma+mF_2}{J-\gamma+mF_2}}.
\label{eq:product_sj}
\end{equation}
Using the Gamma-function identity
\begin{equation}
\prod_{m=1}^{j-1}(m+a)=\frac{\Gamma(j+a)}{\Gamma(1+a)},
\end{equation}
we obtain the closed form
\begin{equation}
d_j=
\left[
\frac{\Gamma\!\left(j+\frac{J+\gamma}{F_2}\right)}{\Gamma\!\left(1+\frac{J+\gamma}{F_2}\right)}
\frac{\Gamma\!\left(1+\frac{J-\gamma}{F_2}\right)}{\Gamma\!\left(j+\frac{J-\gamma}{F_2}\right)}
\right]^{1/2}.
\label{eq:gamma_factor}
\end{equation}
For large $j$, the Gamma function ratio satisfies
\begin{equation}
\frac{\Gamma \left(j+a\right)}{\Gamma \left(j+b\right)} \sim j^{a-b},\qquad j \rightarrow \infty,
\end{equation}
and therefore
\begin{equation}
d_j\sim C j^{\etaloc},
\qquad
\etaloc = \frac{(J+\gamma)-(J-\gamma)}{2F_2}=\frac{\gamma}{F_2},
\label{eq:algebraic_skin}
\end{equation}
where $C$ is a constant independent of $j$. The exponent is independent of the global sign convention $J=\pm1$. This is the first central analytical result.

If $\ket{\phi} = \sum_{j} \phi_j \ket{j}$ is a right eigenstate of $\widetilde{H}$, and $\ket{\psi^R}=\sum_{j} \psi_j^R \ket{j}$ and $\ket{\psi^L} = \sum_{j} \psi_j^L \ket{j}$ are the right and left eigenvectors of the original Hamiltonian $H$, respectively, then they satisfy the following relations:
\begin{equation}
\ket{\psi^R}=D\ket{\phi},
\qquad
\ket{\psi^L}=D^{-1}\ket{\phi}.
\label{eq:left_right_phi}
\end{equation}
The transformed profile is therefore dressed by opposite algebraic factors in the right and left sectors. The exponential skin factor of a uniform nonreciprocal chain is replaced by a much softer algebraic accumulation. The physical reason becomes transparent after taking the logarithm of Eq.~\eqref{eq:s_ratio},
\begin{equation}
\ln\frac{d_{j+1}}{d_j}
=\frac{1}{2}\ln\frac{J+\gamma+jF_2}{J-\gamma+jF_2}
\approx \frac{\gamma}{jF_2}+\mathcal O(j^{-2}).
\label{eq:local_log_increment}
\end{equation}
The local gauge increment decays as $1/j$, so summing over bonds gives $\ln d_j\sim (\gamma/F_2)\ln j$ rather than a quantity proportional to $j$. 

The algebraic skin factor immediately yields a finite-size measure of the residual boundary bias. Summing Eq.~\eqref{eq:local_log_increment} across the whole chain gives
\begin{equation}
\ln\frac{d_N}{d_1}
= \sum_{j=1}^{N-1}\ln\frac{d_{j+1}}{d_j}
= \frac{\gamma}{F_2}\ln N + \mathcal O(1).
\end{equation}
This suggests the finite-size screening variable
\begin{equation}
\Xi_N = \frac{\gamma}{F_2}\ln N.
\label{eq:XiN}
\end{equation}
If $|\Xi_N|\ll 1$, the graded hopping has already screened the nonreciprocal bias across the whole sample. If $|\Xi_N|\gg 1$, the residual algebraic skew remains strong enough to organize finite-size observables.

Figure~\ref{fig:algebraic_check} gives a direct numerical verification of this statement. Panel (a) shows the diagonal similarity factor itself for $N=100$, $F_1=0$, $F_2=1$, and $\gamma=0.5$, where the theoretical exponent is $\eta=\gamma/F_2=0.5$. The log-log plot displays an extended power-law window, confirming that the boundary skew is algebraic rather than exponential. Panel (b) plots the local logarithmic increment $\ln(d_{j+1}/d_j)$ against $1/j$. The near-linearity of the data in the asymptotic region directly reflects Eq.~\eqref{eq:local_log_increment}. Together, the two panels anchor the exact similarity transformation in a concrete observable and show that the graded hopping screens the bare nonreciprocal bias in a bond-dependent way.

\section{Unified asymptotic envelopes and finite-size competition scales}
\label{Unified asymptotic envelopes and finite-size competition scales}

The transformed eigenstate $\ket{\phi}$ obeys
\begin{equation}
\tau_j\phi_{j+1}+\tau_{j-1}\phi_{j-1}+F_1 j\,\phi_j = E\phi_j.
\label{eq:transformed_recurrence}
\end{equation}
At large bond index $j$, Eq.~\eqref{eq:tauj} becomes
\begin{equation}
\tau_j = jF_2\sqrt{1+\frac{2J}{jF_2}-\frac{\gamma^2-J^2}{j^2F_2^2}}
= jF_2 + J + \mathcal O(j^{-1}).
\label{eq:tau_asympt}
\end{equation}
Dividing Eq.~\eqref{eq:transformed_recurrence} by $j$ and discarding the subleading terms gives the constant-coefficient asymptotic recurrence
\begin{equation}
F_2\left(\phi_{j+1}+\phi_{j-1}\right)+F_1\phi_j\simeq 0.
\label{eq:constant_coeff_recurrence}
\end{equation}
With the exponential ansatz $\phi_j\sim r^j$, one obtains
\begin{equation}
r^2+\frac{F_1}{F_2}r+1=0.
\label{eq:characteristic}
\end{equation}
This quadratic equation controls the large-position structure of the transformed problem and, through Eq.~\eqref{eq:left_right_phi}, the large-position envelope of the original non-Hermitian eigenstates.

For $|F_1|<2|F_2|$, the roots are unimodular. Writing
\begin{equation}
\frac{F_1}{F_2}=-2\cos q,
\qquad 0<q<\pi,
\end{equation}
we have $r_{\pm}=\ee^{\pm\ii q}$ and therefore
\begin{equation}
\phi_j \sim A\,\ee^{\ii q j}+B\,\ee^{-\ii q j}.
\end{equation}
The original right eigenstate acquires the asymptotic form
\begin{equation}
\psi_j^R\sim j^{\etaloc}\left(A\,\ee^{\ii q j}+B\,\ee^{-\ii q j}\right),
\qquad |F_1|<2|F_2|.
\label{eq:oscillatory_branch}
\end{equation}
This is not an ordinary extended tail. It is an oscillatory asymptotic structure dressed by an algebraic boundary bias.

For $|F_1|>2|F_2|$, the roots are real. Writing
\begin{equation}
\cosh\kappa = \frac{|F_1|}{2|F_2|},
\qquad \kappa>0,
\label{eq:kappa_def}
\end{equation}
we obtain the decaying branch
\begin{equation}
\phi_j\sim \ee^{-\kappa j},
\qquad |F_1|>2|F_2|,
\end{equation}
and therefore
\begin{equation}
\psi_j^R\sim j^{\etaloc}\ee^{-\kappa j},
\qquad |F_1|>2|F_2|.
\label{eq:localized_branch}
\end{equation}
This expression displays the competition between the two localization channels in its cleanest form. The algebraic factor carries the remnant of nonreciprocity, while the exponential factor is the genuine Stark tail.

Exactly at the threshold $|F_1|=2|F_2|$, the two roots coalesce and the transformed recurrence acquires a double-root solution,
\begin{equation}
\phi_j\sim (A+Bj)r_*^j,
\qquad r_*=\pm 1,
\end{equation}
where the sign depends on the sign of $F_1/F_2$. The original right eigenstate therefore becomes
\begin{equation}
\psi_j^R\sim j^{\etaloc}(A+Bj)r_*^j,
\qquad |F_1|=2|F_2|.
\label{eq:critical_branch}
\end{equation}
More explicitly, Eq.~\eqref{eq:constant_coeff_recurrence} can be written as a transfer-matrix problem,
\begin{equation}
\begin{pmatrix}
\phi_{j+1}\\ \phi_j
\end{pmatrix}
=
T
\begin{pmatrix}
\phi_j\\ \phi_{j-1}
\end{pmatrix},
\qquad
T=
\begin{pmatrix}
-F_1/F_2 & -1\\
1 & 0
\end{pmatrix}.
\label{eq:asymptotic_transfer_matrix}
\end{equation}
The eigenvalues of $T$ are precisely the two roots $r_\pm$. At $|F_1|=2|F_2|$, the transfer matrix has a repeated eigenvalue $r_*=-F_1/(2F_2)=\pm1$ and is not diagonalizable. The factor $j$ in Eq.~\eqref{eq:critical_branch} is therefore the Jordan-block factor of the asymptotic recurrence. The threshold is not merely the point where a localization length becomes large. It is the point where the large-position recurrence changes its algebraic structure. This gives a natural analytical explanation for the broad finite-size threshold region around Eq.~\eqref{eq:main_threshold}.

These three branches are the central analytical result of the paper. They condense the three-way competition into one compact expression. The nonreciprocal bias appears only through the algebraic factor $j^{\eta}$. The Stark physics appears only through the transformed recurrence. The graded hopping mediates both effects at the same time.

On the Stark-localized side, the competition between the algebraic and exponential factors produces a second length scale. Using Eq.~\eqref{eq:localized_branch},
\begin{equation}
\ln|\psi_j^R| \simeq \etaloc\ln j - \kappa j + \mathrm{const}.
\end{equation}
For $\etaloc>0$, corresponding to right-edge accumulation in the conventions of the figures, the maximum of this envelope is obtained by treating $j$ as a continuous variable and setting $\partial_j\ln|\psi_j^R|=0$. This gives
\begin{equation}
j_*\simeq \frac{\etaloc}{\kappa}.
\label{eq:jstar_raw}
\end{equation}
For the opposite sign of nonreciprocity the same estimate applies to the opposite edge after replacing $\etaloc$ by $|\etaloc|$. A dimensionless finite-size measure is therefore
\begin{equation}
\Lambda_N = \frac{|\etaloc|}{\kappa N} = \frac{|\gamma|}{F_2\kappa N}.
\label{eq:LambdaN}
\end{equation}
If $\Lambda_N\gg1$, the exponential Stark decay sets in only beyond the sample size, so the state still appears predominantly boundary-skewed. If $\Lambda_N\ll1$, the Stark tail dominates within the sample.

Near the asymptotic threshold we may write
\begin{equation}
\delta = \frac{|F_1|}{2|F_2|}-1.
\label{eq:delta_def}
\end{equation}
On the localized side, $\delta>0$, and Eq.~\eqref{eq:kappa_def} gives
\begin{equation}
\kappa = \operatorname{arcosh}(1+\delta)\simeq \sqrt{2\delta}
\qquad (\delta\ll1).
\label{eq:kappa_near_threshold}
\end{equation}
The transformed localization length is therefore $\xi\sim\kappa^{-1}\sim\delta^{-1/2}$. If this threshold region is controlled by the condition that the asymptotic tail remains marginal across the whole sample, the natural finite-size estimate is $\xi\sim N$, which yields
\begin{equation}
\delta_N\sim N^{-2}.
\label{eq:ridge_width_scaling}
\end{equation}
In the present manuscript we use this relation as a theoretical guide for the finite-size threshold region, while the main numerical figures focus on the threshold location and the corresponding crossover structure. Combining Eq.~\eqref{eq:LambdaN} with Eq.~\eqref{eq:kappa_near_threshold} also yields a more explicit finite-size estimate at nonzero $\gamma$. Requiring $\Lambda_N\sim1$ gives $\kappa\sim |\gamma|/(F_2N)$ and therefore
\begin{equation}
\delta_{N,\gamma}\sim \frac{\gamma^2}{2F_2^2N^2}.
\label{eq:delta_gamma_scaling}
\end{equation}
This estimate should be read as a crossover estimate rather than as a thermodynamic phase boundary. It shows how the residual algebraic skin factor broadens the apparent threshold region at finite size. The present data are not intended as a precision test of Eq.~\eqref{eq:delta_gamma_scaling}, but the estimate explains why the threshold structure sharpens only slowly with increasing $N$ and why it remains visibly $\gamma$ dependent in finite samples.

\section{Joint localization map in the $(\gamma,F_1/F_2)$ plane}
\label{Joint localization map}

The analytical results above motivate a numerical organization in the $(\gamma,F_1/F_2)$ plane at fixed $F_2$. The horizontal axis measures the distance from the Stark threshold, while the vertical axis tunes the strength of the residual algebraic skin skew. This representation places the two finite-size scales, $\Xi_N$ and $\Lambda_N$, on the same plot. We therefore first use single-particle diagnostics to resolve the localization structure and only then examine the many-body dynamics.

\begin{figure*}[t]
\centering
\includegraphics[width=0.90\textwidth]{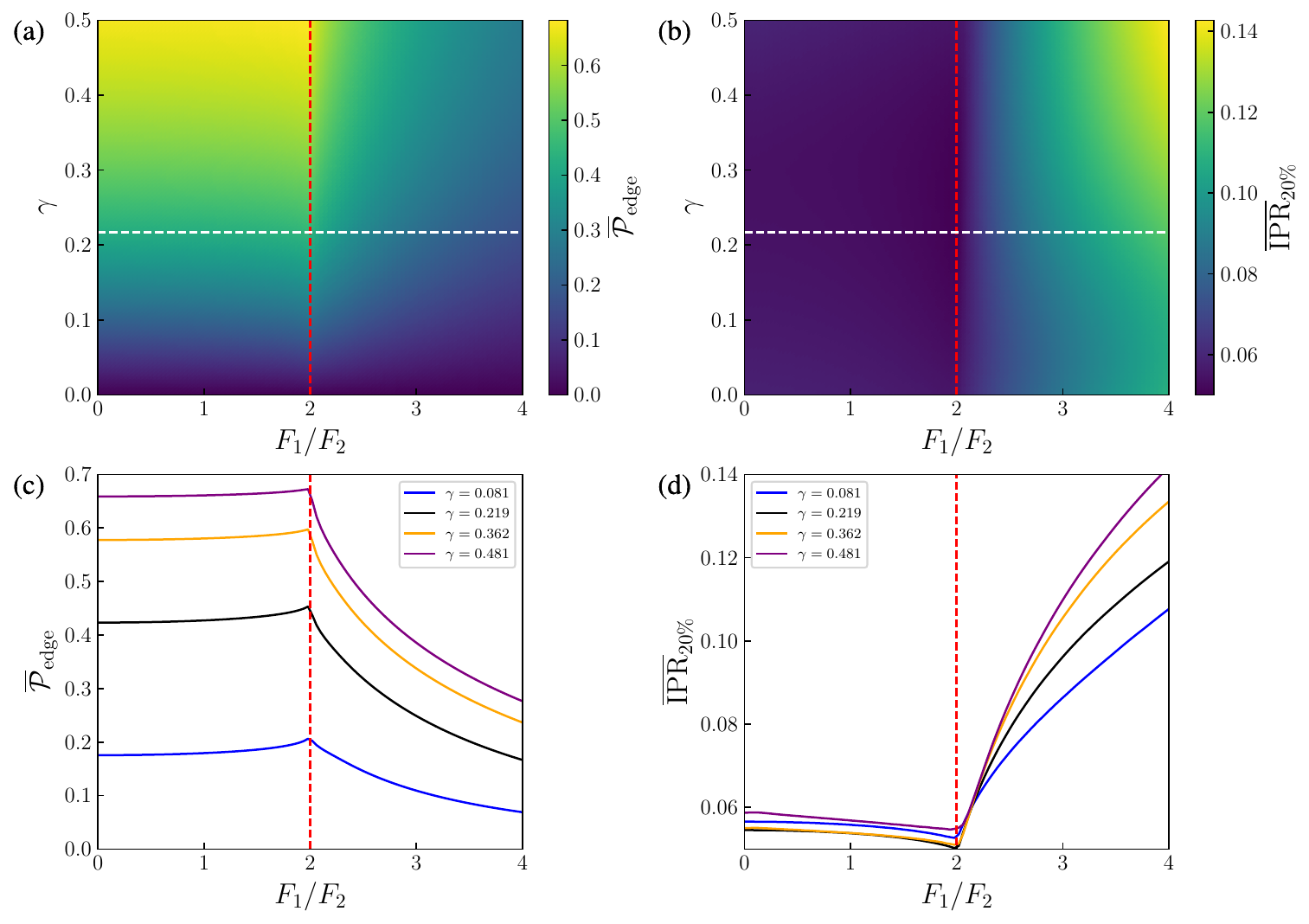}
\caption{Joint localization map at fixed $N=100$, $J=1$, and $F_2=0.20$. Panel (a) shows the mean edge polarization $\overline{\mathcal P}$ of right eigenstates in the $(\gamma,F_1/F_2)$ plane. Panel (b) shows the mean inverse participation ratio built from the most localized 20\% of right eigenstates, $\overline{\IPR}_{20\%}$. The red dashed line marks the asymptotic Stark threshold $F_1/F_2=2$. The white dashed line marks the finite-size screening guide $\Xi_N=(\gamma/F_2)\ln N=5$, corresponding to $\gamma\approx 0.219$. Panels (c) and (d) show representative line cuts at $\gamma=0.081$, $0.219$, $0.362$, and $0.481$. Near the threshold the edge polarization displays a turning point, while the localization measure built from the most localized 20\% develops a minimum and then rises rapidly on the Stark side. The figure directly displays how graded screening of the residual skin bias and the Stark asymptotic threshold reorganize the localization landscape.}
\label{fig:joint_map}
\end{figure*}

Figure~\ref{fig:joint_map} implements this program for $N=100$, $J=1$, and $F_2=0.20$. Panel (a) shows the mean edge polarization of right eigenstates,
\begin{equation}
\overline{\mathcal P}=\frac{1}{N}\sum_{n=1}^{N}\mathcal P_n,
\label{eq:mean_edge_pol}
\end{equation}
while panel (b) shows the mean inverse participation ratio built from the most localized 20\% of right eigenstates,
\begin{equation}
\overline{\IPR}_{20\%}=\frac{1}{N_{20}}\sum_{n\in \mathrm{top}\ 20\%}\IPR_n.
\label{eq:upper_quintile_ipr}
\end{equation}
Here upper quintile denotes the 20\% of right eigenstates with the largest $\IPR_n$ values, so $N_{20}=0.2N$ for the present data set. This second quantity is deliberately more selective than an all-state mean. Its purpose is to isolate the states that control the localized sector and to avoid washing out the crossover by averaging over weakly localized modes. Changing the fraction within a reasonable range gives the same qualitative trend, while the 20\% choice keeps enough states for a smooth finite-size average. The four representative line cuts shown below are taken at $\gamma=0.081$, $0.219$, $0.362$, and $0.481$, so that one of them passes directly through the screening guide $\Xi_N=5$ and the others probe weaker and stronger residual skin bias on the same grid.

Two analytical guides are overlaid on the heat maps. The red dashed line marks the asymptotic Stark threshold $F_1/F_2=2$. The white dashed line marks the finite-size screening condition $\Xi_N=5$, which here corresponds to $\gamma\approx 0.219$. The white line is not a phase boundary. It is a crossover guide that estimates when the residual algebraic skin skew remains strong across the full sample.

The figure assigns a distinct role to the two localization channels. In panel (a), the mean edge polarization grows strongly with $\gamma$ in the weak-Stark region, as expected from the algebraic skin factor. More importantly, for each representative $\gamma$ the line cuts in panel (c) develop a clear turning point near $F_1/F_2=2$. The edge polarization is enhanced as the threshold is approached from the oscillatory side and then decreases once the system moves deeper into the Stark-localized regime. This is the behavior expected from Eq.~\eqref{eq:localized_branch}. Near the threshold the transformed tail remains soft enough for the residual algebraic bias to stay visible. Farther into the Stark sector the exponential factor overwhelms the boundary skew and pulls the center of mass back toward the interior.

The complementary information is carried by panels (b) and (d). The localization measure built from the most localized 20\% of right eigenstates stays relatively flat on the oscillatory side, develops a shallow minimum near $F_1/F_2\simeq2$, and then rises rapidly for $F_1/F_2>2$. This is the direct one-body trace of the Stark tail. The rise is stronger for larger $\gamma$, which is consistent with the competition scale in Eq.~\eqref{eq:LambdaN}. A larger algebraic exponent keeps the profile closer to the boundary before the exponential Stark decay takes over, which sharpens the separation between the threshold region and the more strongly localized sector.

The threshold line therefore enters the figure as a structural feature rather than as an external annotation. The line cuts in panels (c) and (d) do not merely intersect the red guide line. They bend at that location. The numerical map is organized by the analytical scales $\Xi_N$ and $\Lambda_N$ together with the asymptotic threshold $F_1/F_2=2$.

\section{Many-body entanglement growth}
\label{Many-body entanglement growth}

The static localization map suggests a direct dynamical question. If the region around $|F_1|=2|F_2|$ corresponds to the softest large-position tail, then starting from a product state the entanglement growth should be most efficient in the same region. This is the cleanest place where the one-body asymptotics pass over into a genuine many-body observable.

Here we focus on the Hermitian benchmark, $\gamma=0$, where the physical interpretation is especially transparent. The charge-density-wave state of Eq.~\eqref{eq:cdw_state} is convenient because it carries no initial bipartite entanglement. We nevertheless keep the normalized projector formula of Sec.~\ref{Model_and_diagnostics} throughout the calculation so that the dynamical diagnostic is defined in exactly the same way as in the non-Hermitian setting. The early- and intermediate-time growth of $\SL(t)$ is then generated entirely by the hopping dynamics. The physical expectation is simple. For $|F_1|\gg 2|F_2|$, the Stark field produces short-ranged tails and suppresses the transfer of correlations across the cut. For $|F_1|\ll 2|F_2|$, the dynamics is more extended, but the asymptotic recurrence is no longer marginal. Near $|F_1|=2|F_2|$, the transformed recurrence is at its most delicate, so the entanglement growth should be enhanced over an extended time window.

\begin{figure*}[t]
	\centering
	\includegraphics[width=0.90\textwidth]{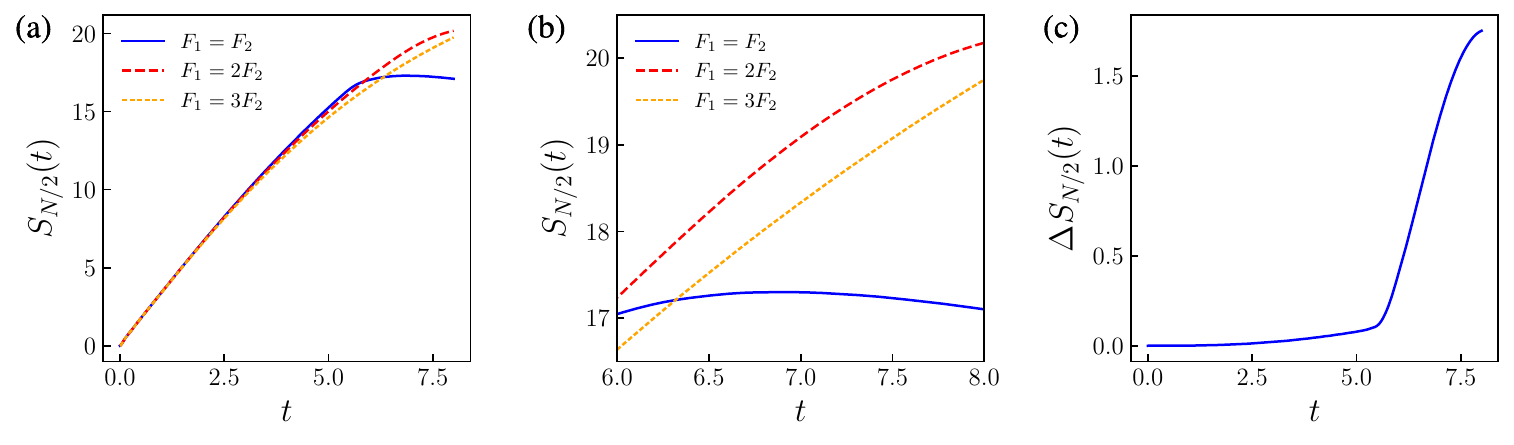}
	\caption{Many-body entanglement growth from the half-filled charge-density-wave state of Eq.~\eqref{eq:cdw_state} in the Hermitian benchmark with $N=120$, $N_p=60$, $J=-1$, $\gamma=0$, and $F_2=0.08$. Panel (a) shows the full time dependence of the half-chain entropy $\SL(t)$ for $F_1/F_2=1$, $2$, and $3$ in the interval $0\le t\le 8$. Panel (b) enlarges the late-time interval $t\in[6,8]$, where the threshold curve $F_1/F_2=2$ clearly lies above the other two. Panel (c) shows the excess entropy $\Delta S(t)$ defined in Eq.~\eqref{eq:deltaS}. The positive late-time excess entropy shows that the asymptotic Stark threshold leaves a direct dynamical trace in the growth of many-body entanglement.}
	\label{fig:manybody_growth}
\end{figure*}

Figure~\ref{fig:manybody_growth} shows that this is indeed what happens. The parameters are $N=120$, $J=-1$, $\gamma=0$, and $F_2=0.08$, and we compare the three representative ratios $F_1/F_2=1$, $2$, and $3$. Because $\gamma=0$, this benchmark contains no skin factor and isolates the Stark part of the asymptotic theory. The choice of $J=-1$ changes short-distance hopping signs but not the leading large-position recurrence, which is set by $F_1/F_2$. Panel (a) shows the full time dependence of the half-chain entropy. The three curves remain close at early times, when the propagation front has not yet felt the asymptotic difference between the three regimes. As time increases, the $F_1/F_2=2$ curve pulls away from its two neighbors.

Panel (b) enlarges the interval $t\in \left[6,8\right]$, where the separation becomes visually clear. The threshold curve sits above both the oscillatory-side curve $F_1/F_2=1$ and the Stark-side curve $F_1/F_2=3$. This ordering persists over an extended late-time window. It is therefore not a fleeting crossing or an artifact of choosing one particular time slice.

To make the comparison sharper, panel (c) plots the excess entropy
\begin{equation}
\Delta S(t)=S_{2F_2}(t)-\frac{1}{2}\left[S_{F_2}(t)+S_{3F_2}(t)\right].
\label{eq:deltaS}
\end{equation}
A positive $\Delta S(t)$ means that the threshold curve lies above the average of the two neighboring curves. In the present data $\Delta S(t)$ remains positive in the late-time window and grows significantly near the end of the plotted interval. This provides a compact dynamical counterpart of the broad threshold region in the one-body problem.

The physical interpretation follows directly from the asymptotic envelopes. Deep in the Stark-dominated region, the exponential factor in Eq.~\eqref{eq:localized_branch} cuts off the propagation of long-range correlations. Deep on the oscillatory side, the state is more extended, but the recurrence is no longer close to its double-root form. Near $|F_1|=2|F_2|$, the transformed tail is at its most delicate, which enhances the transport of correlations across the bipartition and therefore the growth of the many-body entanglement entropy. This dynamical enhancement is not an independent phenomenon. It is the real-time consequence of the same asymptotic threshold that organizes the one-body localization map.

\section{Discussion and conclusion}
\label{Discussion and conclusion}

The main physical message is that graded hopping is not a secondary inhomogeneity added to two familiar localization mechanisms. It is the term that separates their asymptotic roles. After the similarity transformation, nonreciprocity survives as the algebraic factor $j^{\eta}$. The linear potential enters through the transformed recurrence and fixes whether $\phi_j$ is oscillatory, double-root critical, or exponentially decaying. This separation explains why the threshold appears in several different quantities. The edge polarization detects the residual algebraic skin skew. The selective inverse participation ratio detects the onset of the Stark tail. The many-body entropy growth detects the same soft threshold dynamically. The finite-size variables $\Xi_N$ and $\Lambda_N$ give a compact language for these observations and clarify why the data should be viewed as a crossover controlled by asymptotic length scales rather than as a sharp finite-size phase boundary. A useful next step would be a systematic large-$N$ test of the estimates $\delta_N\sim N^{-2}$ and $\delta_{N,\gamma}\sim\gamma^2/(2F_2^2N^2)$, followed by extensions to disorder or interactions where the same separation of gauge dressing and transformed recurrence may persist in renormalized form.

In summary, we have identified a minimal non-Hermitian Stark chain in which skin accumulation, linear-potential localization, and graded hopping can be disentangled analytically. The exact similarity transformation converts exponential skin accumulation into a power-law boundary skew. The transformed large-position recurrence gives the threshold $|F_1|=2|F_2|$. Their product yields the unified envelope $\psi_j^R\sim j^{\eta}\phi_j$, which explains the one-body localization map and the enhancement of half-chain entanglement growth near the same threshold. The agreement between exact asymptotics, finite-size localization diagnostics, and many-body dynamics shows that graded hopping provides a controlled route for reshaping non-Hermitian boundary physics and Stark localization within a single lattice model.

\acknowledgments 

We acknowledge the support of the National Natural Science Foundation of China (Grants No. 12275193, and 11975166).

\appendix

\section{Details of the calculation of many-body entanglement entropy}
\label{Details of the Calculation of Many-Body Entanglement Entropy}

This appendix keeps the detailed derivation of the single-particle method used for the many-body entropy and fixes the notation used in the main text. We distinguish the second-quantized Hamiltonian \(\hat H\) from its one-body matrix \(h\). The occupied orbitals are collected into an \(N\times N_p\) matrix \(U\), where \(N_p\) is the particle number. The overlap matrix in the occupied subspace is denoted by \(M=U^\dagger U\), the single-particle projector by \(P\), and the correlation matrix by \(C\). With the convention used in the main text, \(C_{ij}=\langle c_i^\dagger c_j\rangle\), so \(C=P^T\).

\subsection{Reduction of many-body evolution to a single-particle problem}

The Hamiltonian is quadratic,
\begin{equation}
\hat H=\sum_{\alpha,\beta}c_\alpha^\dagger h_{\alpha\beta}c_\beta,
\label{eq:H_quadric_form}
\end{equation}
where \(h\) is the one-body matrix. The fermion operators satisfy
\begin{align}
\{c_\alpha,c_\beta^\dagger\}&=\delta_{\alpha\beta},\\
\{c_\alpha,c_\beta\}&=\{c_\alpha^\dagger,c_\beta^\dagger\}=0.
\end{align}
A useful commutator is
\begin{equation}
[c_\alpha^\dagger c_\beta,c_m^\dagger]
=c_\alpha^\dagger\{c_\beta,c_m^\dagger\}
-\{c_\alpha^\dagger,c_m^\dagger\}c_\beta
=\delta_{\beta m}c_\alpha^\dagger.
\label{eq:appendix_basic_commutator}
\end{equation}
It follows that
\begin{equation}
[\hat H,c_m^\dagger]
=\sum_{\alpha,\beta}h_{\alpha\beta}[c_\alpha^\dagger c_\beta,c_m^\dagger]
=\sum_\alpha h_{\alpha m}c_\alpha^\dagger.
\label{eq:appendix_first_commutator}
\end{equation}
Repeating the same step gives, by induction,
\begin{equation}
\begin{aligned}
\operatorname{ad}_{\hat H}^n(c_m^\dagger)
& =\sum_\alpha (h^n)_{\alpha m}c_\alpha^\dagger,
\\
\operatorname{ad}_{\hat H}^{n}(B)&=[\hat H,\operatorname{ad}_{\hat H}^{n-1}(B)],\\
\operatorname{ad}_{\hat H}^{0}(B)&=B.
\end{aligned}
\label{eq:appendix_nested_commutator}
\end{equation}
Using the Baker-Campbell-Hausdorff expansion,
\begin{equation}
\ee^{\lambda A}B\ee^{-\lambda A}
=B+\lambda[A,B]+\frac{\lambda^2}{2!}[A,[A,B]]+\cdots,
\end{equation}
with \(A=-\ii t\hat H\), one obtains
\begin{equation}
\ee^{-\ii \hat Ht}c_m^\dagger\ee^{\ii \hat Ht}
=\sum_\alpha \left(\ee^{-\ii ht}\right)_{\alpha m}c_\alpha^\dagger.
\label{eq:appendix_orbital_evolution}
\end{equation}
This identity is the reason why a quadratic many-body evolution can be evaluated through the one-body propagator.

Consider an initial Fock state with occupied sites
\(\mathcal O=\{m_1,m_2,\ldots,m_{N_p}\}\),
\begin{equation}
\ket{\Psi(0)}=c_{m_1}^\dagger c_{m_2}^\dagger\cdots c_{m_{N_p}}^\dagger\ket{\mathrm{vac}}.
\end{equation}
Since \(\hat H\ket{\mathrm{vac}}=0\), Eq.~\eqref{eq:appendix_orbital_evolution} gives
\begin{equation}
\ket{\Psi(t)}
=\prod_{\mu=1}^{N_p}
\left[\sum_{\alpha=1}^{N}
\left(\ee^{-\ii ht}\right)_{\alpha m_\mu}c_\alpha^\dagger\right]
\ket{\mathrm{vac}}.
\label{eq:appendix_single_particle_state}
\end{equation}
Equivalently, if \(U(0)\) is the \(N\times N_p\) matrix whose columns are the occupied basis vectors \(\ket{m_\mu}\), then
\begin{equation}
U(t)=\ee^{-\ii ht}U(0),
\label{eq:appendix_U_t}
\end{equation}
contains the time-evolved occupied orbitals. Equation~\eqref{eq:appendix_single_particle_state} is therefore the Slater determinant built from the columns of \(U(t)\).

\subsection{Normalized correlation matrix for nonorthogonal occupied orbitals}

For a non-Hermitian Hamiltonian, the columns of \(U(t)\) do not in general remain orthonormal. We therefore derive the correlation matrix for a general full-rank orbital matrix \(U\). Define
\begin{equation}
a_\mu^\dagger=\sum_i U_{i\mu}c_i^\dagger,
\qquad
\ket{\Psi_U}=a_1^\dagger a_2^\dagger\cdots a_{N_p}^\dagger\ket{\mathrm{vac}},
\label{eq:appendix_a_def}
\end{equation}
The corresponding annihilation operator is $a_\mu=\sum_i U_{i\mu}^*c_i$, and the occupied-orbital overlap matrix is
\begin{equation}
M_{\mu\nu}=\{a_\mu,a_\nu^\dagger\}
=\sum_i U_{i\mu}^*U_{i\nu}
=(U^\dagger U)_{\mu\nu}.
\label{eq:appendix_overlap}
\end{equation}
The norm of the Slater determinant is the Gram determinant,
\begin{equation}
\braket{\Psi_U}{\Psi_U}=\det M.
\label{eq:appendix_norm}
\end{equation}
This expression follows directly from the fermionic Wick contraction of the string \(a_{N_p}\cdots a_1a_1^\dagger\cdots a_{N_p}^\dagger\). It also shows why \(U\) must have full column rank for the normalized state to be well-defined.

We next compute the normalized correlation matrix. Acting with \(c_j\) on the Slater determinant gives
\begin{equation}
c_j\prod_{\mu=1}^{N_p}a_\mu^\dagger\ket{\mathrm{vac}}
=\sum_{\mu=1}^{N_p}(-1)^{\mu-1}U_{j\mu}
\prod_{\lambda\ne\mu}a_\lambda^\dagger\ket{\mathrm{vac}}.
\label{eq:appendix_cj_action}
\end{equation}
Using the analogous contraction from the left, the numerator of \(\langle c_i^\dagger c_j\rangle\) becomes a sum of minors of \(M\),
\begin{align}
\bra{\Psi_U}c_i^\dagger c_j\ket{\Psi_U}
&=\sum_{\mu,\nu}U_{j\mu}U_{i\nu}^*
(-1)^{\mu+\nu}
\det M^{(\nu,\mu)}.
\label{eq:appendix_minor_sum}
\end{align}
Here \(M^{(\nu,\mu)}\) is the matrix obtained by deleting row \(\nu\) and column \(\mu\) of \(M\). The adjugate matrix satisfies
\begin{equation}
[\operatorname{adj}M]_{\mu\nu}=(-1)^{\mu+\nu}\det M^{(\nu,\mu)},
\qquad
\operatorname{adj}M=(\det M)M^{-1}.
\end{equation}
Dividing Eq.~\eqref{eq:appendix_minor_sum} by the norm \(\det M\), we obtain
\begin{align}
\frac{\bra{\Psi_U}c_i^\dagger c_j\ket{\Psi_U}}{\braket{\Psi_U}{\Psi_U}}
&=\sum_{\mu,\nu}U_{j\mu}(M^{-1})_{\mu\nu}U_{i\nu}^* \\
&=\left[U M^{-1}U^\dagger\right]_{ji}.
\label{eq:appendix_cij_projector_element}
\end{align}
Thus, with the convention \(C_{ij}=\langle c_i^\dagger c_j\rangle\),
\begin{equation}
C=\left[U(U^\dagger U)^{-1}U^\dagger\right]^T.
\label{eq:appendix_projector}
\end{equation}
This is Eq.~\eqref{eq:normalized_projector} of the main text.

It is useful to introduce the matrix without the transpose,
\begin{equation}
P=U(U^\dagger U)^{-1}U^\dagger.
\label{eq:appendix_P_def}
\end{equation}
It is an orthogonal projector onto the occupied single-particle subspace in the standard Hilbert-space inner product. Indeed,
\begin{align}
P^2
&=U M^{-1}U^\dagger U M^{-1}U^\dagger \\
&=U M^{-1}M M^{-1}U^\dagger=P,
\end{align}
and \(P^\dagger=P\). If the columns of \(U\) are orthonormal, \(M=\mathbb I\) and Eq.~\eqref{eq:appendix_projector} reduces to the usual formula, up to the same transpose convention. The entanglement entropy is obtained by restricting \(C\) to the subsystem \(A=\{1,2,\ldots,N/2\}\) and applying Eq.~\eqref{eq:entropy_formula} to the eigenvalues of \(C_A\).

\subsection{QR stabilization}

The formula above is exact, but direct evaluation can become numerically unstable under non-unitary evolution. The columns of \(U(t)\) may develop very different norms and the overlap matrix \(M=U^\dagger U\) may become ill-conditioned. QR stabilization removes this problem without changing the correlation matrix.

The key observation is that the projector is invariant under any invertible change of basis inside the occupied subspace. Let \(B\) be an invertible \(N_p\times N_p\) matrix. Then
\begin{align}
P(UB)
&=UB\left[(UB)^\dagger(UB)\right]^{-1}(UB)^\dagger \\
&=UB\left(B^\dagger U^\dagger U B\right)^{-1}B^\dagger U^\dagger \\
&=UB\left[B^{-1}M^{-1}(B^\dagger)^{-1}\right]B^\dagger U^\dagger \\
&=U M^{-1}U^\dagger=P(U).
\label{eq:appendix_basis_invariance}
\end{align}
Thus the normalized Gaussian projector depends only on the subspace spanned by the occupied orbitals, not on the particular basis used to represent that subspace.

Now take a QR decomposition
\begin{equation}
U=QR,
\label{eq:appendix_QR}
\end{equation}
where \(Q^\dagger Q=\mathbb I_{N_p}\) and \(R\) is invertible and upper triangular. Equation~\eqref{eq:appendix_basis_invariance} gives
\begin{equation}
P(U)=P(QR)=P(Q)=QQ^\dagger.
\label{eq:appendix_Q_projector}
\end{equation}
The same result follows by direct substitution,
\begin{align}
U(U^\dagger U)^{-1}U^\dagger
&=QR(R^\dagger Q^\dagger Q R)^{-1}R^\dagger Q^\dagger \\
&=QR(R^\dagger R)^{-1}R^\dagger Q^\dagger \\
&=QQ^\dagger.
\end{align}
Therefore replacing \(U\) by the orthonormal factor \(Q\) leaves the projector and hence the entropy unchanged.

Finally, we show that QR stabilization can be applied at every time step. Suppose at time \(t_k\) the exact orbital matrix \(U_k\) and the stored stabilized matrix \(Q_k\) span the same occupied subspace. Equivalently, there exists an invertible matrix \(B_k\) such that
\begin{equation}
U_k=Q_kB_k.
\end{equation}
The exact next-step orbitals are
\begin{equation}
U_{k+1}=\ee^{-\ii h\Delta t}U_k
=\ee^{-\ii h\Delta t}Q_k B_k.
\end{equation}
The stabilized algorithm first forms
\begin{equation}
\widetilde U_{k+1}=\ee^{-\ii h\Delta t}Q_k,
\end{equation}
so that \(U_{k+1}=\widetilde U_{k+1}B_k\). Since \(B_k\) is invertible, Eq.~\eqref{eq:appendix_basis_invariance} gives
\begin{equation}
P(U_{k+1})=P(\widetilde U_{k+1}).
\end{equation}
After the QR decomposition \(\widetilde U_{k+1}=Q_{k+1}R_{k+1}\), Eq.~\eqref{eq:appendix_Q_projector} gives
\begin{equation}
P(\widetilde U_{k+1})=P(Q_{k+1}).
\end{equation}
Thus performing QR decomposition after each time step does not alter the normalized correlation matrix. It only replaces an ill-conditioned basis of the same occupied subspace by an orthonormal one. The final entropy is then computed from
\begin{equation}
C_A(t)=\left([P(t)]_{A,A}\right)^T,
\end{equation}
where \([P(t)]_{A,A}\) denotes the submatrix of \(P(t)\) restricted to the sites in the left half of the chain.


\begin{thebibliography}{21}%
	\makeatletter
	\providecommand \@ifxundefined [1]{%
		\@ifx{#1\undefined}
	}%
	\providecommand \@ifnum [1]{%
		\ifnum #1\expandafter \@firstoftwo
		\else \expandafter \@secondoftwo
		\fi
	}%
	\providecommand \@ifx [1]{%
		\ifx #1\expandafter \@firstoftwo
		\else \expandafter \@secondoftwo
		\fi
	}%
	\providecommand \natexlab [1]{#1}%
	\providecommand \enquote  [1]{``#1''}%
	\providecommand \bibnamefont  [1]{#1}%
	\providecommand \bibfnamefont [1]{#1}%
	\providecommand \citenamefont [1]{#1}%
	\providecommand \href@noop [0]{\@secondoftwo}%
	\providecommand \href [0]{\begingroup \@sanitize@url \@href}%
	\providecommand \@href[1]{\@@startlink{#1}\@@href}%
	\providecommand \@@href[1]{\endgroup#1\@@endlink}%
	\providecommand \@sanitize@url [0]{\catcode `\\12\catcode `\$12\catcode
		`\&12\catcode `\#12\catcode `\^12\catcode `\_12\catcode `\%12\relax}%
	\providecommand \@@startlink[1]{}%
	\providecommand \@@endlink[0]{}%
	\providecommand \url  [0]{\begingroup\@sanitize@url \@url }%
	\providecommand \@url [1]{\endgroup\@href {#1}{\urlprefix }}%
	\providecommand \urlprefix  [0]{URL }%
	\providecommand \Eprint [0]{\href }%
	\providecommand \doibase [0]{https://doi.org/}%
	\providecommand \selectlanguage [0]{\@gobble}%
	\providecommand \bibinfo  [0]{\@secondoftwo}%
	\providecommand \bibfield  [0]{\@secondoftwo}%
	\providecommand \translation [1]{[#1]}%
	\providecommand \BibitemOpen [0]{}%
	\providecommand \bibitemStop [0]{}%
	\providecommand \bibitemNoStop [0]{.\EOS\space}%
	\providecommand \EOS [0]{\spacefactor3000\relax}%
	\providecommand \BibitemShut  [1]{\csname bibitem#1\endcsname}%
	\let\auto@bib@innerbib\@empty
	\bibitem [{\citenamefont {Hatano}\ and\ \citenamefont
		{Nelson}(1996)}]{HatanoNelson1996}%
	\BibitemOpen
	\bibfield  {author} {\bibinfo {author} {\bibfnamefont {N.}~\bibnamefont
			{Hatano}}\ and\ \bibinfo {author} {\bibfnamefont {D.~R.}\ \bibnamefont
			{Nelson}},\ }\bibfield  {title} {\bibinfo {title} {Localization transitions
			in non-hermitian quantum mechanics},\ }\href
	{https://doi.org/10.1103/PhysRevLett.77.570} {\bibfield  {journal} {\bibinfo
			{journal} {Phys. Rev. Lett.}\ }\textbf {\bibinfo {volume} {77}},\ \bibinfo
		{pages} {570} (\bibinfo {year} {1996})}\BibitemShut {NoStop}%
	\bibitem [{\citenamefont {Yokomizo}\ and\ \citenamefont
		{Murakami}(2019{\natexlab{a}})}]{YokomizoMurakami2019}%
	\BibitemOpen
	\bibfield  {author} {\bibinfo {author} {\bibfnamefont {K.}~\bibnamefont
			{Yokomizo}}\ and\ \bibinfo {author} {\bibfnamefont {S.}~\bibnamefont
			{Murakami}},\ }\bibfield  {title} {\bibinfo {title} {Non-bloch band theory of
			non-hermitian systems},\ }\href
	{https://doi.org/10.1103/PhysRevLett.123.066404} {\bibfield  {journal}
		{\bibinfo  {journal} {Phys. Rev. Lett.}\ }\textbf {\bibinfo {volume} {123}},\
		\bibinfo {pages} {066404} (\bibinfo {year} {2019}{\natexlab{a}})}\BibitemShut
	{NoStop}%
	\bibitem [{\citenamefont {Okuma}\ \emph {et~al.}(2020)\citenamefont {Okuma},
		\citenamefont {Kawabata}, \citenamefont {Shiozaki},\ and\ \citenamefont
		{Sato}}]{Okuma2020}%
	\BibitemOpen
	\bibfield  {author} {\bibinfo {author} {\bibfnamefont {N.}~\bibnamefont
			{Okuma}}, \bibinfo {author} {\bibfnamefont {K.}~\bibnamefont {Kawabata}},
		\bibinfo {author} {\bibfnamefont {K.}~\bibnamefont {Shiozaki}},\ and\
		\bibinfo {author} {\bibfnamefont {M.}~\bibnamefont {Sato}},\ }\bibfield
	{title} {\bibinfo {title} {Topological origin of non-hermitian skin
			effects},\ }\href {https://doi.org/10.1103/PhysRevLett.124.086801} {\bibfield
		{journal} {\bibinfo  {journal} {Phys. Rev. Lett.}\ }\textbf {\bibinfo
			{volume} {124}},\ \bibinfo {pages} {086801} (\bibinfo {year}
		{2020})}\BibitemShut {NoStop}%
	\bibitem [{\citenamefont {Yokomizo}\ and\ \citenamefont
		{Murakami}(2019{\natexlab{b}})}]{PhysRevLett.123.066404}%
	\BibitemOpen
	\bibfield  {author} {\bibinfo {author} {\bibfnamefont {K.}~\bibnamefont
			{Yokomizo}}\ and\ \bibinfo {author} {\bibfnamefont {S.}~\bibnamefont
			{Murakami}},\ }\bibfield  {title} {\bibinfo {title} {Non-bloch band theory of
			non-hermitian systems},\ }\href
	{https://doi.org/10.1103/PhysRevLett.123.066404} {\bibfield  {journal}
		{\bibinfo  {journal} {Phys. Rev. Lett.}\ }\textbf {\bibinfo {volume} {123}},\
		\bibinfo {pages} {066404} (\bibinfo {year} {2019}{\natexlab{b}})}\BibitemShut
	{NoStop}%
	\bibitem [{\citenamefont {Yao}\ and\ \citenamefont
		{Wang}(2018)}]{PhysRevLett.121.086803}%
	\BibitemOpen
	\bibfield  {author} {\bibinfo {author} {\bibfnamefont {S.}~\bibnamefont
			{Yao}}\ and\ \bibinfo {author} {\bibfnamefont {Z.}~\bibnamefont {Wang}},\
	}\bibfield  {title} {\bibinfo {title} {Edge states and topological invariants
			of non-hermitian systems},\ }\href
	{https://doi.org/10.1103/PhysRevLett.121.086803} {\bibfield  {journal}
		{\bibinfo  {journal} {Phys. Rev. Lett.}\ }\textbf {\bibinfo {volume} {121}},\
		\bibinfo {pages} {086803} (\bibinfo {year} {2018})}\BibitemShut {NoStop}%
	\bibitem [{\citenamefont {Xiao}\ \emph {et~al.}(2020)\citenamefont {Xiao},
		\citenamefont {Deng}, \citenamefont {Wang} \emph {et~al.}}]{Xiao2020}%
	\BibitemOpen
	\bibfield  {author} {\bibinfo {author} {\bibfnamefont {L.}~\bibnamefont
			{Xiao}}, \bibinfo {author} {\bibfnamefont {T.}~\bibnamefont {Deng}}, \bibinfo
		{author} {\bibfnamefont {K.}~\bibnamefont {Wang}}, \emph {et~al.},\
	}\bibfield  {title} {\bibinfo {title} {Non-{Hermitian} bulk–boundary
			correspondence in quantum dynamics},\ }\href
	{https://doi.org/10.1038/s41567-020-0836-6} {\bibfield  {journal} {\bibinfo
			{journal} {Nat. Phys.}\ }\textbf {\bibinfo {volume} {16}},\ \bibinfo {pages}
		{761} (\bibinfo {year} {2020})}\BibitemShut {NoStop}%
	\bibitem [{\citenamefont {Kawabata}\ \emph {et~al.}(2023)\citenamefont
		{Kawabata}, \citenamefont {Numasawa},\ and\ \citenamefont
		{Ryu}}]{Kawabata2023}%
	\BibitemOpen
	\bibfield  {author} {\bibinfo {author} {\bibfnamefont {K.}~\bibnamefont
			{Kawabata}}, \bibinfo {author} {\bibfnamefont {T.}~\bibnamefont {Numasawa}},\
		and\ \bibinfo {author} {\bibfnamefont {S.}~\bibnamefont {Ryu}},\ }\bibfield
	{title} {\bibinfo {title} {Entanglement phase transition induced by the
			non-hermitian skin effect},\ }\href
	{https://doi.org/10.1103/PhysRevX.13.021007} {\bibfield  {journal} {\bibinfo
			{journal} {Phys. Rev. X}\ }\textbf {\bibinfo {volume} {13}},\ \bibinfo
		{pages} {021007} (\bibinfo {year} {2023})}\BibitemShut {NoStop}%
	\bibitem [{\citenamefont {Zhang}\ \emph {et~al.}(2021)\citenamefont {Zhang},
		\citenamefont {Tian}, \citenamefont {Jiang}, \citenamefont {Lu},\ and\
		\citenamefont {Chen}}]{Zhang2021HigherOrderNHSE}%
	\BibitemOpen
	\bibfield  {author} {\bibinfo {author} {\bibfnamefont {X.}~\bibnamefont
			{Zhang}}, \bibinfo {author} {\bibfnamefont {Y.}~\bibnamefont {Tian}},
		\bibinfo {author} {\bibfnamefont {J.-H.}\ \bibnamefont {Jiang}}, \bibinfo
		{author} {\bibfnamefont {M.-H.}\ \bibnamefont {Lu}},\ and\ \bibinfo {author}
		{\bibfnamefont {Y.-F.}\ \bibnamefont {Chen}},\ }\bibfield  {title} {\bibinfo
		{title} {Observation of higher-order non-{{Hermitian}} skin effect},\ }\href
	{https://doi.org/10.1038/s41467-021-25716-y} {\bibfield  {journal} {\bibinfo
			{journal} {Nat Commun}\ }\textbf {\bibinfo {volume} {12}},\ \bibinfo {pages}
		{5377} (\bibinfo {year} {2021})}\BibitemShut {NoStop}%
	\bibitem [{\citenamefont {Li}\ \emph {et~al.}(2024{\natexlab{a}})\citenamefont
		{Li}, \citenamefont {Wang}, \citenamefont {Wang} \emph
		{et~al.}}]{Li2024DynamicNHSE}%
	\BibitemOpen
	\bibfield  {author} {\bibinfo {author} {\bibfnamefont {Z.}~\bibnamefont
			{Li}}, \bibinfo {author} {\bibfnamefont {L.-W.}\ \bibnamefont {Wang}},
		\bibinfo {author} {\bibfnamefont {X.}~\bibnamefont {Wang}}, \emph {et~al.},\
	}\bibfield  {title} {\bibinfo {title} {Observation of dynamic
			non-{{Hermitian}} skin effects},\ }\href
	{https://doi.org/10.1038/s41467-024-50776-1} {\bibfield  {journal} {\bibinfo
			{journal} {Nat Commun}\ }\textbf {\bibinfo {volume} {15}},\ \bibinfo {pages}
		{6544} (\bibinfo {year} {2024}{\natexlab{a}})}\BibitemShut {NoStop}%
	\bibitem [{\citenamefont {Ben~Dahan}\ \emph {et~al.}(1996)\citenamefont
		{Ben~Dahan}, \citenamefont {Peik}, \citenamefont {Reichel}, \citenamefont
		{Castin},\ and\ \citenamefont {Salomon}}]{BenDahan1996}%
	\BibitemOpen
	\bibfield  {author} {\bibinfo {author} {\bibfnamefont {M.}~\bibnamefont
			{Ben~Dahan}}, \bibinfo {author} {\bibfnamefont {E.}~\bibnamefont {Peik}},
		\bibinfo {author} {\bibfnamefont {J.}~\bibnamefont {Reichel}}, \bibinfo
		{author} {\bibfnamefont {Y.}~\bibnamefont {Castin}},\ and\ \bibinfo {author}
		{\bibfnamefont {C.}~\bibnamefont {Salomon}},\ }\bibfield  {title} {\bibinfo
		{title} {Bloch oscillations of atoms in an optical potential},\ }\href
	{https://doi.org/10.1103/PhysRevLett.76.4508} {\bibfield  {journal} {\bibinfo
			{journal} {Phys. Rev. Lett.}\ }\textbf {\bibinfo {volume} {76}},\ \bibinfo
		{pages} {4508} (\bibinfo {year} {1996})}\BibitemShut {NoStop}%
	\bibitem [{\citenamefont {Wilkinson}\ \emph {et~al.}(1996)\citenamefont
		{Wilkinson}, \citenamefont {Bharucha}, \citenamefont {Madison}, \citenamefont
		{Niu},\ and\ \citenamefont {Raizen}}]{Wilkinson1996}%
	\BibitemOpen
	\bibfield  {author} {\bibinfo {author} {\bibfnamefont {S.~R.}\ \bibnamefont
			{Wilkinson}}, \bibinfo {author} {\bibfnamefont {C.~F.}\ \bibnamefont
			{Bharucha}}, \bibinfo {author} {\bibfnamefont {K.~W.}\ \bibnamefont
			{Madison}}, \bibinfo {author} {\bibfnamefont {Q.}~\bibnamefont {Niu}},\ and\
		\bibinfo {author} {\bibfnamefont {M.~G.}\ \bibnamefont {Raizen}},\ }\bibfield
	{title} {\bibinfo {title} {Observation of atomic wannier-stark ladders in an
			accelerating optical potential},\ }\href
	{https://doi.org/10.1103/PhysRevLett.76.4512} {\bibfield  {journal} {\bibinfo
			{journal} {Phys. Rev. Lett.}\ }\textbf {\bibinfo {volume} {76}},\ \bibinfo
		{pages} {4512} (\bibinfo {year} {1996})}\BibitemShut {NoStop}%
	\bibitem [{\citenamefont {Chakrabarty}\ and\ \citenamefont
		{Datta}(2025)}]{PhysRevB.111.174202}%
	\BibitemOpen
	\bibfield  {author} {\bibinfo {author} {\bibfnamefont {A.}~\bibnamefont
			{Chakrabarty}}\ and\ \bibinfo {author} {\bibfnamefont {S.}~\bibnamefont
			{Datta}},\ }\bibfield  {title} {\bibinfo {title} {Fate of wannier-stark
			localization and skin effect in periodically driven non-hermitian
			quasiperiodic lattices},\ }\href
	{https://doi.org/10.1103/PhysRevB.111.174202} {\bibfield  {journal} {\bibinfo
			{journal} {Phys. Rev. B}\ }\textbf {\bibinfo {volume} {111}},\ \bibinfo
		{pages} {174202} (\bibinfo {year} {2025})}\BibitemShut {NoStop}%
	\bibitem [{\citenamefont {Gl\"uck}\ \emph {et~al.}(2000)\citenamefont
		{Gl\"uck}, \citenamefont {Hankel}, \citenamefont {Kolovsky},\ and\
		\citenamefont {Korsch}}]{PhysRevA.61.061402}%
	\BibitemOpen
	\bibfield  {author} {\bibinfo {author} {\bibfnamefont {M.}~\bibnamefont
			{Gl\"uck}}, \bibinfo {author} {\bibfnamefont {M.}~\bibnamefont {Hankel}},
		\bibinfo {author} {\bibfnamefont {A.~R.}\ \bibnamefont {Kolovsky}},\ and\
		\bibinfo {author} {\bibfnamefont {H.~J.}\ \bibnamefont {Korsch}},\ }\bibfield
	{title} {\bibinfo {title} {Wannier-stark ladders in driven optical
			lattices},\ }\href {https://doi.org/10.1103/PhysRevA.61.061402} {\bibfield
		{journal} {\bibinfo  {journal} {Phys. Rev. A}\ }\textbf {\bibinfo {volume}
			{61}},\ \bibinfo {pages} {061402} (\bibinfo {year} {2000})}\BibitemShut
	{NoStop}%
	\bibitem [{\citenamefont {Guo}\ \emph {et~al.}(2021)\citenamefont {Guo},
		\citenamefont {Ge}, \citenamefont {Li} \emph {et~al.}}]{Guo2021WSL}%
	\BibitemOpen
	\bibfield  {author} {\bibinfo {author} {\bibfnamefont {X.~Y.}\ \bibnamefont
			{Guo}}, \bibinfo {author} {\bibfnamefont {Z.~Y.}\ \bibnamefont {Ge}},
		\bibinfo {author} {\bibfnamefont {H.}~\bibnamefont {Li}}, \emph {et~al.},\
	}\bibfield  {title} {\bibinfo {title} {Observation of {Bloch} oscillations
			and {Wannier-Stark} localization on a superconducting quantum processor},\
	}\href {https://doi.org/10.1038/s41534-021-00385-3} {\bibfield  {journal}
		{\bibinfo  {journal} {npj Quantum Inf}\ }\textbf {\bibinfo {volume} {7}},\
		\bibinfo {pages} {51} (\bibinfo {year} {2021})}\BibitemShut {NoStop}%
	\bibitem [{\citenamefont {Karamlou}\ \emph {et~al.}(2022)\citenamefont
		{Karamlou}, \citenamefont {Braum\"{u}ller}, \citenamefont {Yanay} \emph
		{et~al.}}]{Karamlou2022}%
	\BibitemOpen
	\bibfield  {author} {\bibinfo {author} {\bibfnamefont {A.~H.}\ \bibnamefont
			{Karamlou}}, \bibinfo {author} {\bibfnamefont {J.}~\bibnamefont
			{Braum\"{u}ller}}, \bibinfo {author} {\bibfnamefont {Y.}~\bibnamefont
			{Yanay}}, \emph {et~al.},\ }\bibfield  {title} {\bibinfo {title} {Quantum
			transport and localization in {1d} and {2d} tight-binding lattices},\ }\href
	{https://doi.org/10.1038/s41534-022-00528-0} {\bibfield  {journal} {\bibinfo
			{journal} {npj Quantum Inf}\ }\textbf {\bibinfo {volume} {8}},\ \bibinfo
		{pages} {35} (\bibinfo {year} {2022})}\BibitemShut {NoStop}%
	\bibitem [{\citenamefont {Peng}\ \emph {et~al.}(2022)\citenamefont {Peng},
		\citenamefont {Jie}, \citenamefont {Yu},\ and\ \citenamefont
		{Wang}}]{PhysRevB.106.L161402}%
	\BibitemOpen
	\bibfield  {author} {\bibinfo {author} {\bibfnamefont {Y.}~\bibnamefont
			{Peng}}, \bibinfo {author} {\bibfnamefont {J.}~\bibnamefont {Jie}}, \bibinfo
		{author} {\bibfnamefont {D.}~\bibnamefont {Yu}},\ and\ \bibinfo {author}
		{\bibfnamefont {Y.}~\bibnamefont {Wang}},\ }\bibfield  {title} {\bibinfo
		{title} {Manipulating the non-hermitian skin effect via electric fields},\
	}\href {https://doi.org/10.1103/PhysRevB.106.L161402} {\bibfield  {journal}
		{\bibinfo  {journal} {Phys. Rev. B}\ }\textbf {\bibinfo {volume} {106}},\
		\bibinfo {pages} {L161402} (\bibinfo {year} {2022})}\BibitemShut {NoStop}%
	\bibitem [{\citenamefont {Zheng}\ \emph {et~al.}(2024)\citenamefont {Zheng},
		\citenamefont {Jalali~Mehrabad}, \citenamefont {Vannucci} \emph
		{et~al.}}]{Zheng2024}%
	\BibitemOpen
	\bibfield  {author} {\bibinfo {author} {\bibfnamefont {X.}~\bibnamefont
			{Zheng}}, \bibinfo {author} {\bibfnamefont {M.}~\bibnamefont
			{Jalali~Mehrabad}}, \bibinfo {author} {\bibfnamefont {J.}~\bibnamefont
			{Vannucci}}, \emph {et~al.},\ }\bibfield  {title} {\bibinfo {title} {Dynamic
			control of {2D} non-{Hermitian} photonic corner skin modes in synthetic
			dimensions},\ }\href {https://doi.org/10.1038/s41467-024-55236-4} {\bibfield
		{journal} {\bibinfo  {journal} {Nat Commun}\ }\textbf {\bibinfo {volume}
			{15}},\ \bibinfo {pages} {10881} (\bibinfo {year} {2024})}\BibitemShut
	{NoStop}%
	\bibitem [{\citenamefont {Ghosh}\ and\ \citenamefont
		{Roy}()}]{ghoshExploringTopologicalLocalization2025}%
	\BibitemOpen
	\bibfield  {author} {\bibinfo {author} {\bibfnamefont {S.}~\bibnamefont
			{Ghosh}}\ and\ \bibinfo {author} {\bibfnamefont {S.}~\bibnamefont {Roy}},\
	}\bibfield  {title} {\bibinfo {title} {Exploring topological and localization
			phenomena in ssh chains under generalized aah modulation: A computational
			approach},\ }\href {https://arxiv.org/abs/2506.10195} {\bibinfo  {journal}
		{arXiv:2506.10195}\ }\BibitemShut {NoStop}%
	\bibitem [{\citenamefont {Li}\ \emph {et~al.}(2024{\natexlab{b}})\citenamefont
		{Li}, \citenamefont {Wan},\ and\ \citenamefont
		{Zhong}}]{PhysRevB.110.094310}%
	\BibitemOpen
	\bibfield  {journal} {  }\bibfield  {author} {\bibinfo {author} {\bibfnamefont
			{H.-Z.}\ \bibnamefont {Li}}, \bibinfo {author} {\bibfnamefont
			{M.}~\bibnamefont {Wan}},\ and\ \bibinfo {author} {\bibfnamefont {J.-X.}\
			\bibnamefont {Zhong}},\ }\bibfield  {title} {\bibinfo {title} {Fate of
			non-hermitian free fermions with wannier-stark ladder},\ }\href
	{https://doi.org/10.1103/PhysRevB.110.094310} {\bibfield  {journal} {\bibinfo
			{journal} {Phys. Rev. B}\ }\textbf {\bibinfo {volume} {110}},\ \bibinfo
		{pages} {094310} (\bibinfo {year} {2024}{\natexlab{b}})}\BibitemShut
	{NoStop}%
	\bibitem [{\citenamefont {Szyniszewski}\ \emph {et~al.}(2023)\citenamefont
		{Szyniszewski}, \citenamefont {Lunt},\ and\ \citenamefont
		{Pal}}]{PhysRevB.108.165126}%
	\BibitemOpen
	\bibfield  {author} {\bibinfo {author} {\bibfnamefont {M.}~\bibnamefont
			{Szyniszewski}}, \bibinfo {author} {\bibfnamefont {O.}~\bibnamefont {Lunt}},\
		and\ \bibinfo {author} {\bibfnamefont {A.}~\bibnamefont {Pal}},\ }\bibfield
	{title} {\bibinfo {title} {Disordered monitored free fermions},\ }\href
	{https://doi.org/10.1103/PhysRevB.108.165126} {\bibfield  {journal} {\bibinfo
			{journal} {Phys. Rev. B}\ }\textbf {\bibinfo {volume} {108}},\ \bibinfo
		{pages} {165126} (\bibinfo {year} {2023})}\BibitemShut {NoStop}%
	\bibitem [{\citenamefont {Li}\ \emph {et~al.}()\citenamefont {Li},
		\citenamefont {Liu},\ and\ \citenamefont
		{Xu}}]{liDisorderInducedEntanglementPhase2023a}%
	\BibitemOpen
	\bibfield  {author} {\bibinfo {author} {\bibfnamefont {K.}~\bibnamefont
			{Li}}, \bibinfo {author} {\bibfnamefont {Z.-C.}\ \bibnamefont {Liu}},\ and\
		\bibinfo {author} {\bibfnamefont {Y.}~\bibnamefont {Xu}},\ }\bibfield
	{title} {\bibinfo {title} {Disorder-induced entanglement phase transitions in
			non-hermitian systems with skin effects},\ }\href
	{https://arxiv.org/abs/2305.12342} {\bibinfo  {journal} {arXiv:2305.12342}\
	}\BibitemShut {NoStop}%
\end{thebibliography}
%

\end{document}